\documentclass[a4paper,fleqn,usenatbib]{mnras}

\usepackage{newtxtext,newtxmath}

\usepackage[T1]{fontenc}
\usepackage{ae,aecompl}

\usepackage{graphicx}	
\usepackage{amsmath}	
\usepackage{amssymb}	
\usepackage{hyperref}
\usepackage{multirow}

\usepackage{tikz}

\usepackage{pifont}
\newcommand{\cmark}{\ding{51}}%

\usepackage{etoolbox}
\makeatletter
\patchcmd\@combinedblfloats{\box\@outputbox}{\unvbox\@outputbox}{}{%
   \errmessage{\noexpand\@combinedblfloats could not be patched}%
}%
 \makeatother

\title[Variability in the quasar J1100-0053]{A new physical interpretation of optical and infrared variability in quasars}

\author[N.P. Ross et al.]
{Nicholas~P.~Ross$^{1}$\thanks{E-mail: npross@roe.ac.uk},    
K. E. Saavik Ford$^{2,3,4}$,  Matthew Graham$^{5}$,  Barry McKernan$^{2,3,4}$,  
\newauthor Daniel Stern$^{6}$, Aaron M. Meisner$^{7,8}$, Roberto J. Assef$^{9}$, 
Arjun Dey$^{10}$, Andrew J. Drake$^{11}$, 
\newauthor Hyunsung D. Jun$^{12}$, Dustin Lang$^{13,14,15}$
\\
$^{1}$Institute for Astronomy, University of Edinburgh, Royal Observatory, Blackford Hill, Edinburgh EH9 3HJ, United Kingdom \\
$^{2}$Department of Science, BMCC, City University of New York, New York, NY 10007, USA \\
$^{3}$Department of Astrophysics, Rose Center for Earth and Space, American Museum of Natural History, Central Park West at 79th Street, NY 10024, USA \\
$^{4}$Graduate Center, City University of New York, 365 5th Avenue, New York, NY 10016, USA\\
$^{5}$Cahill Center for Astronomy and Astrophysics, California Institute of Technology, Mail Code 249/17, 1200 E California Blvd, Pasadena CA 91125, USA\\
$^{6}$Jet Propulsion Laboratory, California Institute of Technology, 4800 Oak Grove Drive, Mail Stop 169-221, Pasadena, CA 91109, USA \\
$^{7}$Lawrence Berkeley National Laboratory, 1 Cyclotron Road, Berkeley, CA 92420, U.S.A. \\
$^{8}$Berkeley Center for Cosmological Physics, Berkeley, CA 94720, USA\\
$^{9}$N\'ucleo de Astronom\'ia de la Facultad de Ingenier\'ia y Ciencias, Universidad Diego Portales, Av. Ej\'ercito Libertador 441, Santiago, Chile \\
$^{10}$National Optical Astronomy Observatory, 950 N. Cherry Ave, Tucson, AZ 85719, USA \\
$^{11}$Center for Advanced Computing Research, California Institute of Technology, 1200 E California Blvd, Pasadena CA 91125, USA \\
$^{12}$School of Physics, Korea Institute for Advanced Study, 85 Hoegiro, Dongdaemun-gu, Seoul 02455, Korea\\
$^{13}$Dunlap Institute, University of Toronto, Toronto, ON M5S 3H4, Canada \\
$^{14}$Department of Astronomy \& Astrophysics, University of Toronto, Toronto, ON M5S 3H4, Canada \\
$^{15}$Perimeter Institute for Theoretical Physics, Waterloo, ON N2L 2Y5, Canada\\
}

\date{Accepted XXX. Received YYY; in original form ZZZ}
\pubyear{2018}

\begin{document}
\label{firstpage}
\pagerange{\pageref{firstpage}--\pageref{lastpage}}
\maketitle

\begin{abstract}
Changing-look quasars are a recently identified class of active galaxies in which the strong UV continuum and/or broad optical hydrogen emission lines associated with unobscured quasars either appear or disappear on timescales of months to years.  The physical processes responsible for this behaviour are still debated, but changes in the black hole accretion rate or accretion disk structure appear more likely than changes in obscuration. Here we report on four epochs of spectroscopy of SDSS J110057.70-005304.5, a quasar at a redshift of $z=0.378$ whose UV continuum and broad hydrogen emission lines have faded, and then returned over the past $\approx$20 years. The change in this quasar was initially identified in the infrared, and an archival spectrum from 2010 shows an intermediate phase of the transition during which the flux below rest-frame $\approx$3400\AA\ has decreased by close to an order of magnitude. This combination is unique compared to previously published examples of changing-look quasars, and is best explained by dramatic changes in the innermost regions of the accretion disk. The optical continuum has been rising since mid-2016, leading to a prediction of a rise in hydrogen emission line flux in the next year. Increases in the infrared flux are beginning to follow, delayed by a $\sim$3 year observed timescale. If our model is confirmed, the physics of changing-look quasars are governed by processes at the innermost stable circular orbit (ISCO) around the black hole, and the structure of the innermost disk. The easily identifiable and monitored changing-look quasars would then provide a new probe and laboratory of the nuclear central engine.
\end{abstract}


\section{Introduction}
The Shakura-Sunyaev $\alpha-$disk model \citep{SS73} has long been
used to describe the basic properties of the optically thick,
geometrically thin accretion disks expected to orbit the supermassive
black holes at the nuclei of quasars. This accretion disk is thought
to be the origin of thermal continuum emission observed in the
rest-frame ultraviolet and optical. The thermal emission seen in the
infrared spectrum of quasars is believed to originate from molecular
dust outside the accretion disk and traditional broadline region
(BLR). Thus, the IR flux is directly proportional to the emission from
the disk, reprocessed by the dusty reservoir and delayed by the
light-travel time between the two \citep[see e.g.,][for
reviews]{Antonucci1993, Perlman2008, Lasota2016}. As such, the thermal
accretion disk photons are the seeds for both the X-ray emission --
due to Compton-upscattering in the corona
\citep[e.g.,][]{Begelman1983, Risaliti2009, Lusso2017} and the thermal
mid-IR emission from the torus.

The $\alpha$-disk model assumes that the disk is geometrically thin
(i.e., $h/r \ll 1$ where $h/r$ is the disk aspect ratio) and that
angular momentum is transported by a kinematic viscosity, $\nu$,
parametrized by \citet{SS73} as 
\begin{equation}
\nu = \alpha c_{\rm s} h 
\end{equation}
where
$c_{\rm s}$ is the local mean sound speed in the disk and $h$ is the
scale-height perpendicular to the disk plane. The thermal emission
need not be, but often is, treated as a superposition of blackbodies
at varying annuli with an effective temperature
dependence\footnote{This r$^{-3/4}$ temperature dependence is not
specific to the $\alpha$-parametrization, and is independent from the
nature of the viscosity, provided that the disk is geometrically thin,
steady, with heat dissipation and angular momentum transport caused
by the same, local mechanism.} going as $T(r) \propto r^{-3/4}$.

Given the size scales and temperatures associated with supermassive
black holes, a substantial fraction of the bolometric luminosity
should be in the form of UV photons -- the so-called ``Big Blue Bump''
\citep{Shields1978, Malkan_Sargent1982}. For the optically thick, UV
emitting disk to accrete onto the black hole, substantial angular
momentum must be lost.  The kinematic viscosity of the plasma,
$\alpha$, seems the likely mechanism that transports angular momentum
outward.  This viscosity is likely due to magnetorotational
instability \citep[MRI; ][]{Balbus_Hawley1991} with additional
contributions to turbulence from the effects of objects embedded in
the disk \citep[e.g.,][]{McKernan2014}.

However, as has long been established \citep[e.g., ][]{Alloin1985} and
recently re-visited \citep[e.g., ][]{LaMassa2015, Runnoe2016,
MacLeod2016, Ruan2016, Rumbaugh2017, Yang2017, Lawrence2018}, the
observation of even slowly varying Balmer emission lines in quasars
strongly suggests that if a thermal accretion disk does indeed
contribute substantially to the ionizing or optical continuum, then it
can not be in quasi-steady state equilibrium. The variations must be
due to more chaotic disturbances or instabilities in the disk that
propagate at considerably higher speeds than the radial accretion flow
and possibly as fast as the orbital velocity.

Furthermore, as e.g., \citet{Koratkar_Blaes1999} and
\citet{Sirko_Goodman2003} among others point out, the observed
spectral energy distributions (SEDs) of typical quasars differ
markedly from classical $\alpha$-disk theoretical predictions
\citep[][]{SS73, Pringle1981} with a typical observed quasar SED flat
in $\lambda F_{\lambda}$ over several decades in wavelength
\citep{Elvis1994, Richards2006b}. Also, real AGN disks seem to be
cooler \citep[e.g., ][]{Lawrence2012} and larger
\citep[e.g.,][]{Pooley2007, Morgan2010, Morgan2012, Mosquera2011} than
the $\alpha$-disk model predicts. Furthermore, the $\alpha$-disk is an
ad hoc parameterization of disk viscosity and does not permit
predictions of global changes from local perturbations
\citep{King2012}.

Nevertheless, in this paper, we utilize the mathematically simple
$\alpha$-disk model as a framework and departure point for our own
disk models. Here, we build on previous work \citep{Sirko_Goodman2003,
Zimmerman2005, Hameury2009} and introduce a phenomenological model
which does allow changes across the accretion disk and, crucially,
makes predictions which can be observed in the SED.

Changing-look quasars (CLQs) are luminous active galaxies in which the
strong UV continuum and/or broad optical hydrogen emission lines
associated with unobscured quasars either appear or disappear on
timescales of months to years. CLQs have traditionally been discovered
by looking for large, $| \Delta m | >1$ magnitude changes in the
optical light curves of quasars or galaxies. In contrast, we have
taken advantage of the ongoing mid-IR Near-Earth Object Wide-Field
Infrared Survey Explorer Reactivation mission \citep[NEOWISE-R;
][]{Mainzer2014}, supplemented with the optical Dark Energy Camera
Legacy Survey (DECaLS\footnote{{\tt legacysurvey.org/decamls/}}), in
order to discover new changing-look quasars.  While previous efforts
have used the 1-year baseline of the WISE mission to identify
changing-look quasars \citep[e.g.,][]{Assef2018, Stern2018}, our
investigation is the first to extend this selection to the infrared
using NEOWISE-R mission data. We have identified a sample of Sloan
Digital Sky Survey (SDSS) quasars that show significant changes in
their IR flux over the course of a few years. Importantly, our IR
light curves enable us to set limits on SED changes due to
obscuration.

In this article we present the $z=0.378$ quasar SDSS
J110057.70-005304.5 (hereafter J1100-0053).  J1100-0053 was a known
quasar we identified as interesting due to its IR light curve. We have
spectral observations for J1100-0053 showing a transition in the blue
continuum into a `dim state' where the rest-frame UV flux is
suppressed, and then returning to a state with a strong blue
continuum. The model we present invokes changes at the innermost
stable circular orbit (ISCO, defined as $r_{\textsc{ISCO}}={\frac
{6\,GM}{c^{2}}}$ in the Schwarzschild metric) to be the triggering
event for substantial changes in the wider accretion disk, including
major structural changes out to $\approx$225$r_{g}$ (where $r_{\rm g}$ is the
gravitational radius; $r_{\rm g}=\frac{GM}{c^2}$). Such a model
explains changes in the broad emission lines, as well as the optical
and IR light curves.

This paper is organised as follows. In Section 2, we describe our
sample selection, catalogs and observationa data sets.  In Section 3,
we present various theoretical models and discuss if and how each
describes and explains the data.  We conclude in Section 4.  We report
all magnitudes on the AB zero-point system \citep{Oke_Gunn1983,
Fukugita1996} unless otherwise stated explicitly. For the WISE bands,
$m_{\rm AB} = m_{\rm Vega} + m$ where $m = (2.699, 3.339)$ for WISE W1
at 3.4$\mu$m and WISE W2 at 4.6$\mu$m, respectively
\citep{Cutri2011}. We choose the cosmological parameters
$\Omega_{\Lambda} = 0.7$, $\Omega_{\rm M} = 0.3$, and $h = 0.7$ in
order to be consistent with \citet{Shen2011}.

\begin{figure*}
  \centering
  \includegraphics[width=16.00cm, height=11.00cm, trim=0.0cm 0.0cm 0.0cm 0.0cm, clip]
  {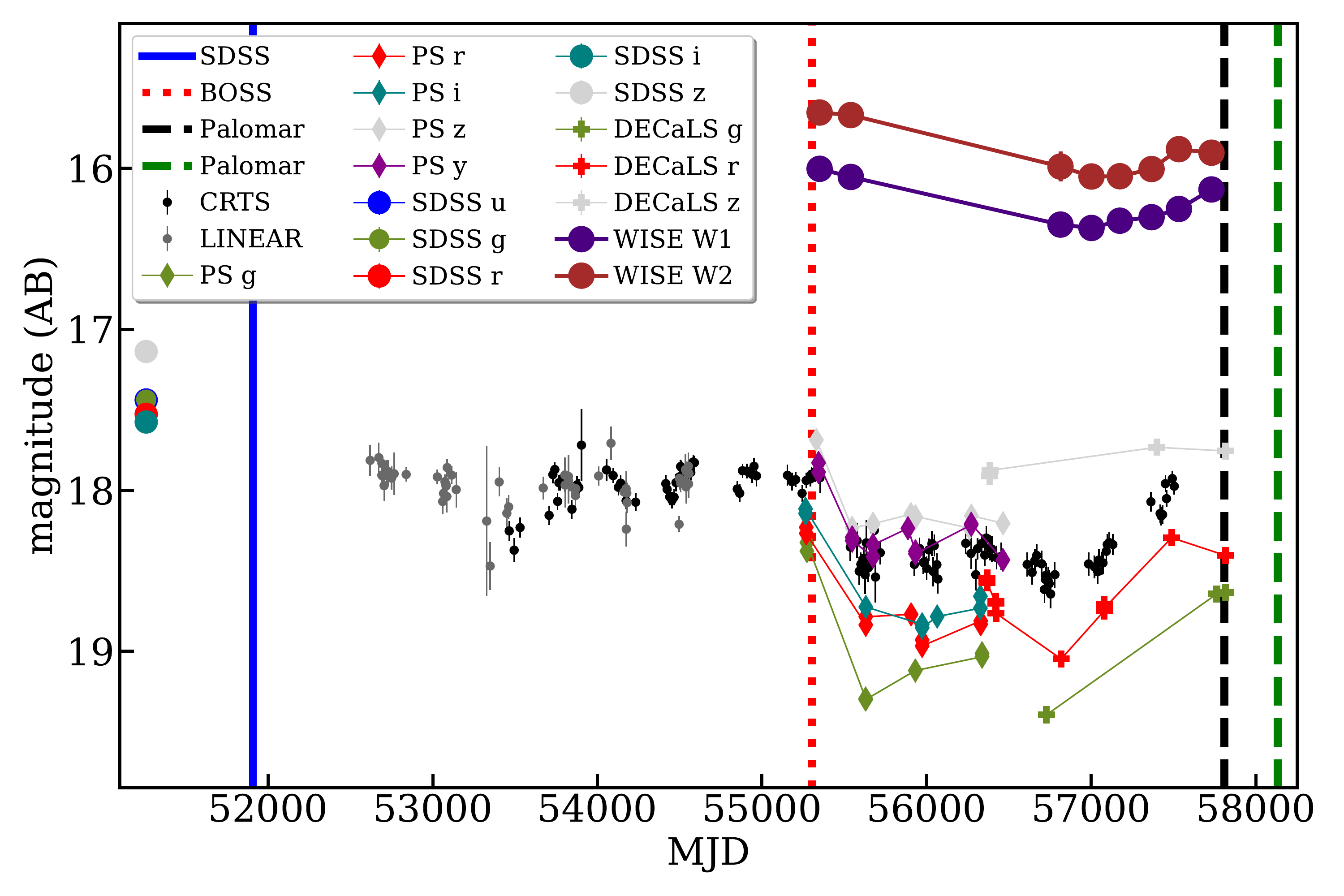}
  \caption[]{
    Multi-wavelength light curve of J1100-0053, including optical data
    from LINEAR, CRTS, SDSS, PanSTARRS and DECaLS, and mid-IR data from
    the WISE satellite.  The four vertical lines illustrate the four
    epochs of optical spectra presented in Figure 2.  J1100-0053 was
    flagged for further study due to the IR fading observed by WISE.  Note
    that the optical emission has been recovering over the past few years,
    with the IR emission beginning to show similar behvaiour.}
  \label{fig:J110057_LC_CRTS}
\end{figure*}
\section{Target Selection and Observations}  

\subsection{Selection in SDSS and NEOWISE-R of J1100-0053}
We started by matching the SDSS Data Release 7 \cite[DR7Q;
][]{Schneider2007} and the SDSS-III Baryon Oscillation Spectroscopic
Survey (BOSS) Data Release 12 quasar catalogues \cite[DR12Q;
][]{Paris2017} to the NEOWISE-R IR data. We use data from the
beginning of the WISE mission \citep[2010 January; ][]{Wright2010}
through the thrid-year of NEOWISE-R operations \citep[2016 December;
][]{Mainzer2011}. The WISE scan pattern leads to coverage of the
full-sky approximately once every six months (a ``sky pass''), but the
satellite was placed in hibernation in 2011 February and then
reactivated in 2013 October. Hence, our light curves have a cadence of
6 months with a 32 month sampling gap.

The W1/W2 light curves for $\sim$200,000 SDSS and BOSS spectroscopic
quasars were obtained by performing forced photometry at the locations
of DECam-detected optical sources \citep{Lang2014, Meisner2017a,
Meisner2017b}. This forced photometry was performed on time-resolved
coadds \citep{Lang2014}, each of which represents a stack of $\sim$12
exposures. The coaddition removes the possibility of probing
variability on $\lesssim$1 day time scales, but pushes $\approx$1.4
magnitudes deeper than individual exposures while removing virtually
all single-exposure artifacts (e.g. cosmic rays and satellites).

Approximately $\sim$30,000 of the SDSS/BOSS quasars with W1/W2
light-curves available are `IR-bright', in that they are above both
the W1 and W2 single exposure thresholds and therefore detected at
very high significance in the coadds. For this ensemble of objects,
the typical variation in each quasar's measured (W1-W2) color is 0.06
magnitudes.  This includes statistical and systematic errors which are
expected to contribute variations at the few hundredths of a magnitude
level. The typical measured single-band scatter is 0.07 magnitudes in
each of W1 and W2.

We undertook a search for outliers relative to these
trends. Specifically, we selected objects with the following
characteristics:
\begin{itemize}
  \item Monotonic variation in both W1 and W2 flux.
  \item W1 flux and W2 flux Pearson correlation coefficient $r \geq0.9$.
  \item $>$0.5 mag peak-to-peak variation in either W1 or W2.
\end{itemize}
This yields a sample of 248 sources. 31 of these are assumed to be
blazars due to the presence of Faint Images of the Radio Sky at
Twenty-Centimeters \citep[FIRST; ][]{Becker1995} radio counterparts,
and we discount them for further analyses. Another 22 objects are
outside the FIRST footprint, leaving 195 quasars in our IR-variable
sample, with no potential FIRST counterparts detected within 30''. 

Although aperture photometry and DECaLS forced photometry
\citep{Lang2014, Meisner2017a, Meisner2017b} are available, J1100-0053
is significantly above the single-exposure detection limit so it is
valid to obtain photometry from the publicly released W1/W2 Level 1b
(L1b) single-exposure images at the NASA/IPAC Infrared Science Archive
(\href{http://irsa.ipac.caltech.edu/}{IRSA}).  Upon querying the
combined the WISE All-Sky, WISE Post-Cryo and NEOWISE-R databases, we
have 101 measurements in 8 sky passes spanning nearly 2400 days.

Links to all our data, catalogs and analysis can be found
online at: \href{https://github.com/d80b2t}{{\tt github.com/d80b2t}}.

\subsection{Optical Imaging}
Figure~\ref{fig:J110057_LC_CRTS} presents the light curve of SDSS
J1100-0053.  J1100-0053 was first detected in the National
Geographic Society-Palomar Observatory Sky Survey \cite[NGS-POSS;
][]{Abell1959, Minkowski_Abell1963book} in 1955 April. It is
catalogued in the SuperCOSMOS Science Archive
\citep[\href{http://ssa.roe.ac.uk/}{SSA}; ][]{Hambly2001_I,
Hambly2001_II} and due to its equatorial position was also observed by
the UK Schmidt Telescope \cite[UKST; ][]{Cannon1975,
Cannon1979book}. Querying the SSA returns {\it gCorMag} and {\it
sCorMag} which are the magnitudes assuming the object is either a
galaxy or star, respectively. We use the {\it sCorMag} values as is
appropriate for an image with flux dominated by the point-like AGN;
the {\it sCorMag} magnitudes are calibrated in the Vega system. For
J1100-0053 we find the magnitudes are: 18.10 mag in the blue UK-J
filter from MJD 45440.47 (1983 April 16); 17.49 mag in the red POSS-I
`E'-filter from MJD 35214.22 (1955 April 17); 17.92 mag in the red
UK-R filter from MJD 46521.47 (1986 April 01) and 17.71 mag in the
UK-I filter from MJD 47273.49 (1988 April 22). J1100-0053 is not in
the Digital Access to a Sky Century @ Harvard
(DASCH\footnote{http://dasch.rc.fas.harvard.edu/project.php}).

J1100-0053 was imaged by the Sloan Digital Sky Survey (SDSS) in the
$u$, $g$, $r$, $i$ and $z$-bands in 1999 March, and more recently by
the Dark Energy Camera Legacy Survey (DECaLS) where there are 4, 13
and 4 exposures in the $g$, $r$ and $z$-bands, respectively, in the
DECaLS Data Release 3 \citep[DR3; ][]{Dey2018}. The $g$-band
observations span $\approx$3 years ($56727 \leq g_{\rm MJD} \leq
57816$), while the $r$- and $z$-band observations span $\approx$4
years ($56367 \leq r_{\rm MJD} \leq 57814$ and $56383 \leq z_{\rm MJD}
\leq 57815$).

Along with WISE IR data, optical data from the SDSS, Catalina
Real-time Transient Survey \citep[CRTS;][]{Drake2009, Mahabal2011},
the Lincoln Near-Earth Asteroid Research \citep[LINEAR; ][]{Sesar2011}
program and the Panoramic Survey Telescope and Rapid Response System
\citep[PanSTARRS;][]{Kaiser2010, Stubbs2010, Tonry2012, Magnier2013}
are also available, and presented in Fig.~\ref{fig:J110057_LC_CRTS}.

\begin{figure*}
  \centering
  \includegraphics[width=17.00cm, trim=0.0cm 0.0cm 0.0cm 0.0cm, clip]
  {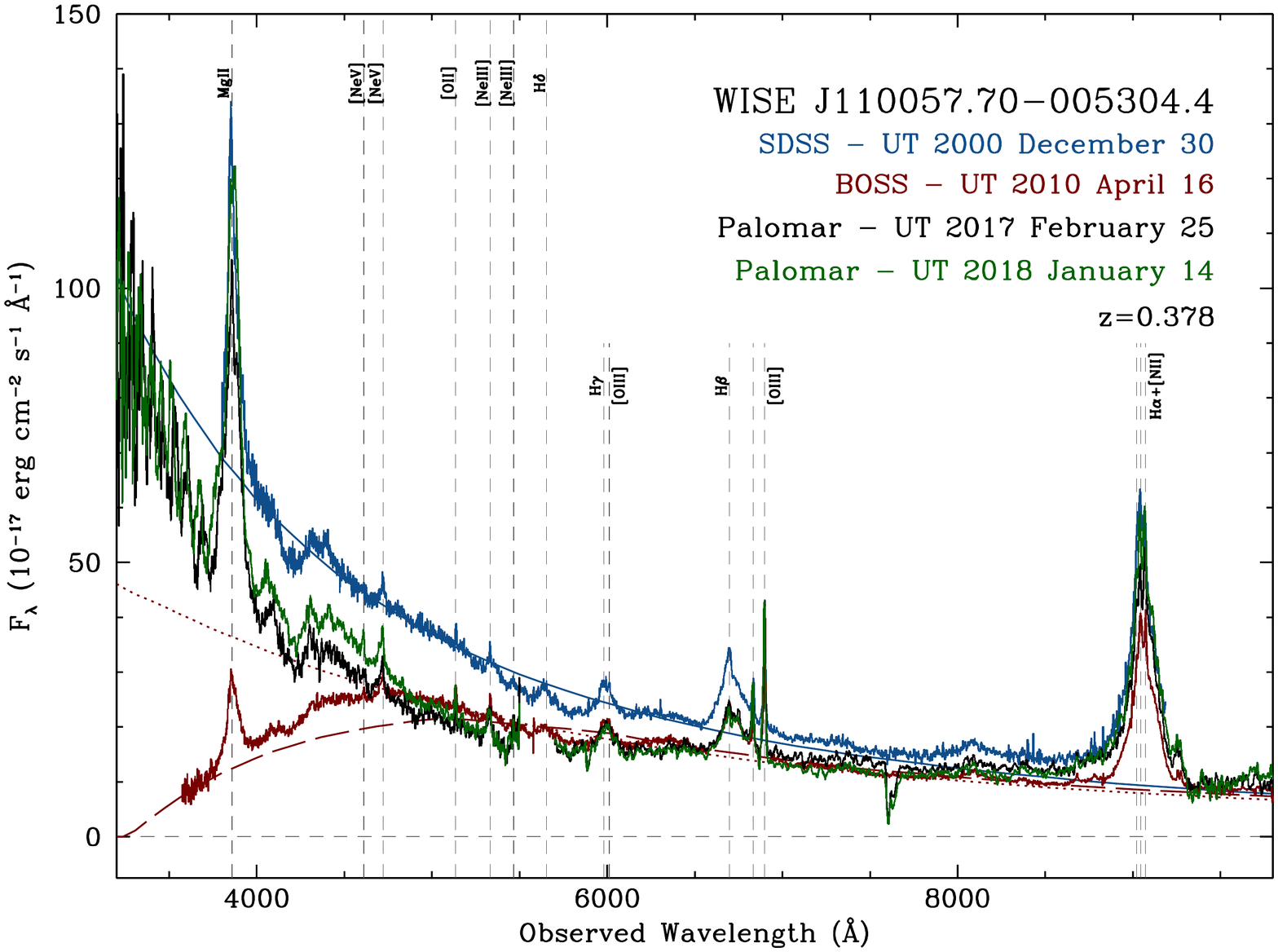}
\vspace{-16pt}
  \caption[]{
    Optical spectra of J1100-0053 obtained on MJD 51908 (blue; SDSS),
    55302 (red; BOSS), 57809 (black; Palomar) and 58132 (green;
    Palomar). Spectra have been renormalized to maintain a constant
    [O\,{\sc iii}]\ luminosity. Over the past two decades, the UV
    continuum and broad lines have changed significantly for this quasar.
    In particular, the 2nd-epoch BOSS spectrum from 2010 shows the rare
    occurrence of a temporary collapse of the UV continuum.  
    The smooth lines show three simple thermal accretion disk models
    of the continuum.  As discussed in Section 3, torque at the ISCO,
    possibly due to magnetic fields threading the inner disk and plunging
    region, heats the inner disk, causing it to puff up and become more UV
    luminous. The solid blue line shows an inflated disk with non-zero
    torque at the ISCO. The dotted red line shows a modified zero-torque
    model where the thermal disk emission interior to 40 $r_{\rm g}$
    is suppressed by a factor of 10.  The red long dashed line is a zero
    torque at the ISCO model multiplied by an absorption law adapted from
    \citet{Guo2016}.
}
  \label{fig:J110057_spectra}
\end{figure*}
\subsection{Additional Multiwavelength Data for J1100-0053}
J1100-0053 was observed by R\"{o}ntgensatellit (ROSAT) and appears in
the All-Sky Survey Bright Source Catalogue \citep[RASS-BSC;
][]{Appenzeller1998, Voges1999} as 2RXS J110058.1-005259 with 27.00
counts (count error 6.14) and a count rate = 0.06$\pm$0.01
counts~s$^{-1}$ \citep{Boller2016}. The NASA/IPAC Extragalactic
Database (NED\footnote{https://ned.ipac.caltech.edu/}) gives
J1100-0053 as having a flux of $1.27\pm0.28 \times10^{-12}$ erg cm$^{-2}$
s$^{-1}$ in the 0.1-2.4 keV range (unabsorbed). J1100-0053 is not in
either the {\it Chandra} or XMM-{\it Newton} archives but is detected
by the Galaxy Evolution Explorer \citep[GALEX; ][]{Martin2005,
Morrissey2007} and has reported flux densities of $19.29\pm0.12$ mag in
the far-UV and $18.89\pm0.05$ mag in the near-UV. As noted above,
there is no radio counterpart within 30 arcsec in the FIRST survey,
i.e. at 21cm. None of the {\it Hubble Space Telescope}, {\it Spitzer
Space Telescope} or {\it Kepler} missions have observed J1100-0053.
It is also not in the Hyper Suprime-Cam (HSC)
\href{https://hsc-release.mtk.nao.ac.jp/doc/}{Data Release 1}
\citep{Aihara2017} footprint.

\subsection{Spectroscopy}
\subsubsection{SDSS and BOSS Spectroscopy}

Figure~\ref{fig:J110057_spectra} shows four optical spectra of
J1100-0053. J1100-0053 satisfied a number of spectroscopic targeting
flags, making it a quasar target in SDSS \citep{Richards2002}. An SDSS
spectrum was obtained on MJD 51908 (SDSS Plate 277, Fiber 212) and the
spectrum of a $z=0.378$ quasar was catalogued in the SDSS Early Data
Release \citep{Stoughton2002, Schneider2002}. 

The second epoch spectrum is from the SDSS-III Baryon Oscillation
Spectroscopic Survey \citep[BOSS; ][]{Dawson2013} on MJD 55302 and
shows the downturn at $\lesssim$4300\AA\ (observed). SDSS-III BOSS
actively vetoed previously known $z<2$ quasars \citep{Ross2012}, but
due to J1100-0053 being selected as an ancillary target \citep[via a
white dwarf program;][]{Kepler2015, Kepler2016} a second spectral
epoch was obtained. Due to a design tradeoff to improve throughput in
the Ly$\alpha$-forest of quasar spectra in BOSS, quasar targets were
subject to spectrophotometric calibration errors
\citep{Margala2016}. These are introduced primarily due to offsets in
fiber-hole positioning between quasar targets and spectrophotometric
standard stars. However, since J1100-0053 was {\it not} a BOSS quasar
{\it target}, it is not subject to this ``blue offset''. J1100-0053
has no pipeline flag suggesting the spectrum was compromised during
data taking. We checked the calibration of BOSS Plate 3836 that
observed J1100-0053 and confirmed that the data were high-SNR and that
the behaviour in the blue spectrum was not due to the instrument,
telescope or data reduction.

One significant aspect of the BOSS spectrum is the strong, broad
Mg\,{\sc ii}\ emission line atop a fading red continuum.  Mg\,{\sc
ii}\ emission being prominent in objects without otherwise strong blue
continuum is very unusual, but not unheard of.  For example,
\citet{Roig2014} find a group of objects with strong and broad
Mg\,{\sc ii}\ line emission, but very weak H$\alpha$ and H$\beta$
emission, and undetectably low near-UV AGN continuum flux.

\subsubsection{Palomar Spectroscopy} 
A third epoch spectrum was obtained from the Palomar Hale 5m telescope
using the Double Spectrograph (DBSP) instrument.  Exposures of 600s
and 300s were taken in good conditions on UT 2017 February 25 (MJD
57809). Features to note include the continuum straddling Mg\,{\sc
ii}\ being blue in the 2017 spectrum, as it was for the SDSS spectrum
in 2000, as opposed to red, as it was for the BOSS spectrum in 2010. A
fourth spectral epoch was also taken using the Hale 5m and DBSP on
UT 2018 January 14 (MJD 58132). 

The first-epoch SDSS spectrum shows a typical blue quasar, but the
blue continuum decreases by nearly a factor of ten in flux in the
second epoch BOSS spectrum taken 10 years later. The blue continuum
then returns in the third epoch spectrum taken another 7 years later,
albeit at a diminished level relative to the initial spectrum. The
final spectrum obtained, approximately 1 year late, is similar to the
third epoch spectrum.

The SDSS and BOSS spectroscopy is performed with 3'' and 2'' fibers,
respectively. The Palomar observations are done with 1.5 arcsec
slits. We note that the 2017 Palomar data was photometric, while 2018
Palomar data was not.  Both Palomar spectra were flux calibrated using
the sensitivity function calculated from the photometric night (in
2017), and then the spectra were scaled to have approximately matching
[O\,{\sc iii}] line strengths. As such, in the following, we note that
the larger [O\,{\sc iii}] luminosities obtained with the fiber spectra
as opposed to the slit spectra is consistent with [O\,{\sc iii}] being
obtained through wider apertures and the line being somewhat extended.

\subsubsection{Emission Line Fitting} 
We fit the SDSS spectrum for J1100-0053 with the models provided by
QSFit\footnote{http://qsfit.inaf.it/}\citep{Calderone2017} and require
three components in order to replicate the observed asymmetric
H$\beta$ emission line profile: a narrow FWHM=1000 km s$^{-1}$
component, and two broad (11897$\pm$556 and 4594$\pm$341 km s$^{-1}$)
components. The two broad components overlap and hence QSFit consider
them as a single line, whose profile is given by the sum of the two
Gaussians. The final FWHM of 7415$\pm$499 km s$^{-1}$ is calculated on
the latter profile.  Thus, we suggest that J1100-0053, at least from
its SDSS epoch, could be an ``anomalous H$\beta$ quasar'' as described
by \citet{Steinhardt_Silverman2013}. Those authors suggest that
$\sim$1/4 of all SDSS quasars at $z < 0.8$ for which H$\beta$ is
well-measured is best described by not one, but two well-centered
broad components, resulting in the ``anomalous'' broadened H$\beta$
lines.

\begin{table*}
 \centering
  \begin{tabular}{l r r r r}
    \hline \hline 
    Spectrum                             &	SDSS  &	BOSS    &	Palomar  & Palomar \\
    MJD                                     &    51908 &  55302  &    57809   &  58132   \\          
    \hline   
    FWHM H$\beta$ (broad)     &  5907$\pm$141      & 6102$\pm$91    & 6684$\pm$180  &7657$\pm$239  \\
    log (BH mass / $M_{\odot}$) & 8.89$\pm$0.12       & 8.82$\pm0.12$  & 8.56$\pm$0.12  & 8.76$\pm$0.12  \\
   log L( 5100\AA\ )                 & 44.78$\pm$0.01   &44.56$\pm$0.01  &43.96$\pm$0.04 & 44.11$\pm$0.02 \\
  log L(H$\beta$, broad)        & 43.05$\pm$0.02    & 42.76$\pm$0.02 &42.29$\pm$0.02   & 42.44$\pm$0.03 \\
    log L(H$\beta$, narrow)    & 41.26$\pm$0.61     & 41.02$\pm$0.61 & 40.41$\pm$0.61 & 40.67$\pm$0.61\\
   log L(H$\beta$, total)         & 43.06$\pm$0.02    &42.76$\pm$0.02 &42.294$\pm$0.02 & 42.44$\pm$0.03\\
log  L([O\,{\sc iii}])                & 42.20$\pm$0.04   &42.29$\pm$0.01 & 41.61$\pm$0.11& 41.86$\pm$0.03 \\
    H$\beta$  / [O\,{\sc iii}]         &  7.18 & 3.07   & 4.89 & 3.79  \\
   5100 \AA\  / [O\,{\sc iii}]         &  378 &  203 &  225 & 176  \\
   \hline \hline 
 \end{tabular}
 \caption{Spectral values to the H$\beta$ [O\,{\sc iii}] line complex using the methods
    of \citet{Jun2015a}.  FWHM in km s$^{-1}$ and luminosities are in ergs s$^{-1}$.  
  The final two rows give the ratio of the total H$\beta$ to total [O\,{\sc iii}] luminosities, 
and total 5100 \AA\ continuum to total [O\,{\sc iii}] line luminosities, respectively.} 
\label{tab:Hbeta_details}
\end{table*}

Using the methods presented in \citet{Jun2015a}, we perform further
fits to the H$\beta$ and [O\,{\sc iii}] line complex and the results
are given in Table~\ref{tab:Hbeta_details}.  The total H$\beta$ (with
2 broad+1 narrow components) to total [O\,{\sc iii}] (1 broad+1
narrow, where the narrow width/center is fixed to the Balmer
components) ratio drops by a factor $\times$2.3 times and 1.5, 1.9
times in the third and fourth epochs. The 5100\AA\ continuum to total
[O\,{\sc iii}] ratios also change by $\times$1.9, 1.7 and 2.1 in the
2nd, 3rd, and 4th epochs, indicating that the 5100\AA\ and H$\beta$
luminosities vary consistently. We return to the H$\beta$ line profile
in our modeling discussion in Section 3.

While continuum changes in the rest-frame UV/optical spectra of
quasars are not a new discovery (see e.g., \citealt{Clavel1991}, the
review by \citealt{Ulrich1997} and more recent studies by
\citealt{VandenBerk2004, Pereyra2006, MacLeod2010} and
\citealt{Guo2016b}), the identification of a ``UV collapse'' for
quasars has only recently been noted by \cite{Guo2016}. Those
authors report the first discovery of a UV cutoff quasar, SDSS
J231742.60 +000535.1 (hereafter J2317+0005; redshift $z = 0.321$),
observed spectroscopically by SDSS three times, on UT 2000 September 29, UT 2001
September 25, and UT 2001 October 18. In the case of J2317+0005, a
cycle of UV emission collapse, quasar dimming, and recovery was
observed over the course of just a few weeks. For J1100-0053, the
cycle is far longer; however, the combination of optical and infrared
light curves, as well as observing J1100-0053 at four separate
spectral stages is currently unique. As such, J1100-0053 and
J2317+0005 are now two archetypal objects that any accretion disk
model must predict and explain \citep[e.g.,][]{Lawrence2018}.

In our companion study, \citet{Stern2018} report on a new
changing-look quasar, J1052+1519, identified with the same selection
as J1100-0053, and where the broad H$\beta$ emission has vanished
compared to an archival SDSS spectrum. The physical properties of
J1100-0053 derived from the MJD 51908 spectrum using the methods in
\citet{Shen2011}, are given in Table~\ref{tab:Shen_props}, where we
also give the properties of J2317+0005 \citep{Guo2016} and J1052+1519
\citep{Stern2018} for comparison.

\begin{table*}
 \centering
 \begin{tabular}{l l l l}
  \hline \hline 
 Quantity                                          &     this paper                       &  Guo et al. (2016)              & Stern et al. (2018) \\
 \hline 
    &&\\
    SDSS name                                       &    J110057.70-005304.5    &  J231742.60+000535.1    & J105203.55+151929.5 \\
    R.A. / deg                                        &  165.240463                       &  349.42752075                &   163.01480103  \\
    Declination / deg                            &   -0.884586                        &   +0.093091                    & 15.32488632 \\
   redshift, $z$                                    &   0.3778$\pm$0.0003        & 0.3209$\pm$0.0002       & 0.3022$\pm$0.0008\\
    &&\\ 
    \multirow{3}{*}{SDSS Plate, Fiber, MJD }  &  	*277, 212,   51908    &  *382, 173, 51816         & *2483, 204, 53852 \\
    &                                           & 679, 551, 52177              & \\    
    &                                           & 680, 346, 52200             & \\
    BOSS Plate, Fiber, MJD                 & 	3836, 258, 55302          	    &    --                                & -- \\
    $M_{i}(z=2)$  / mag                          &   -24.48                             & -23.65                           & -22.73 \\
    log $(L_{\rm bol} / {\rm erg s}^{-1}) $  &   45.78$\pm$0.02               & 45.56$\pm$0.004      & 45.07$\pm$0.004 \\
    log $(M_{\rm BH} / M_{\odot})  $           &  8.83$\pm$0.14                & 8.43$\pm$0.03           & 8.46$\pm$0.02 \\
    Eddington ratio  (\%)                        &        7.0                         &  10.7                           &  3.2     \\ 
    &&\\
    \hline \hline 
  \end{tabular}
  \caption{Physical properties of J1100-0053, J2317+0005 and J1052+1519 using the
    methods from \citet{Shen2011}. *This spectrum was used to estimate
    the quantities reported.  We use the regular definition of $L_{\rm
      Edd} = 4 \pi G M m_{\rm p} c /\sigma_{T} =
    1.26\times10^{38}\left (M/M_{\odot} \right )$ erg s$^{-1}$.} 
 \label{tab:Shen_props}
\end{table*}

\section{Phenomenological Modeling} 
In a similar vein to the discussion in \cite{Stern2018}, in this
section we discuss several models with the aim of determining the
physical mechamism(s) driving the light curve and spectral behaviour
of J1100-0053. The explanations come in two broad classes: obscuration
and changes in the accretion disk. Ultimately, we are forced towards
a model of the latter type that combines a cooling front propagating
in the accretion disk along with changes in the disk opacity.

\subsection{Scenario I: Obscuration by an Infalling Cloud}
We explore the possibility that an obscuring cloud, or clouds, cause
the observed light curve and spectral behaviour of J1100-0053. This
explanation is dismissed for the CLQ J0159+0033 in \citet{LaMassa2015}
but is the preferred explanation for J2317+0005 in \citet{Guo2016}.

In this scenario, the obscuring cloud(s) are required to cross the line
of sight. The clouds also need to block most of the inner disk such
that the ionizing radiation could not impact on the BLR or the torus
for a period of months to years, in order to explain both the IR drop and
broadline disappearance. An explanation of why the light curves
`recover' after a period of $\sim$2500 days (observed-frame) is also
required; i.e., why do the light curves not rapidly return to their
original flux levels once the obscuring event is over.

Clouds should not typically infall; they need to lose angular momentum
if they are drawn from a distribution with Keplerian orbits, and even
if they do lose angular momentum, e.g., in a collision with clouds of 
approximately equal mass, they would likely be either destroyed or no
longer coherent. The relevant timescales here are the freefall and
cloud-crushing times. The freefall timescale is 
\begin{equation}
    t_{{\rm ff}}   \sim 100   {\rm yr}  
                     \left( \frac{ r } {0.4\rm{pc} }\right)^{3/2} 
                                            \left(\frac{M}{10^{8}M_{\odot}}\right)^{-1}
\end{equation}
and Kelvin-Helmholtz instabilities would destroy the clouds within the
cloud-crushing time \citep[e.g., ][]{Nagakura2008, Hopkins2013,
Shiokawa2015, Bae2016}, given by
\begin{equation}
    t_{\rm cc} \sim 100{\rm yr} \left(\frac{\rho_{{\rm cloud}}/\rho_{{\rm medium}}} {10^{6}}\right)^{1/2} 
                                            \left(\frac{r_{\rm cloud}}{4 \times 10^{10} \, \rm{km}}\right) 
                                            \left(\frac{v_{\rm rel}}{10^{4} \, \rm{km s^{-1}}}\right)^{-1}.
\label{eqn:T_cloudcrushing}
\end{equation}
Thus, even if clouds did infall, they would end up fragmented, which
should pollute the inner disk.  The dust in the cloud would then be
well inside the dust sublimation radius 
\begin{equation}
    R_{\rm dust} \approx 0.4{\rm pc}\left(\frac{L}{10^{45}\rm{erg s^{-1}}}\right)^{1/2}
                                                   \left(\frac{T_{\rm sub}}{1500\rm{K}}\right)^{2.6}
\end{equation}
and so the dust will be destroyed in the $\sim$100 year free-fall from
the dust-sublimation region. Hence, one can not absorb the UV spectrum
with dust, since it will have been sublimated well before it arrives
at the inner disk.

Given the time interval spanned by the spectroscopic observations
($\approx$7.7 yr), it is physically possible that an obscuring cloud,
just large enough to block our view of the inner disk region could be
in a circular orbit at the sublimation radius; this cloud could then
cross its own diameter in approximately six years depressing the UV
emission. To check this scenario, we use the standard assumptions from
\citet{Whittet1992book} with $\rho_{\rm dust}$ being expressed in
terms of $<A/L>$, extinction per kpc for the ISM:
\begin{equation}
\rho_{\rm dust}= 1.2 \times 10^{-27} \cdot  s \cdot (n^2 + 2/n^2 -1) \cdot <A/L>  
\end{equation}
\citep[eqn. 3.30 in][]{Whittet1992book},  
where $s$ is the  specific density (e.g. 2500~kg~m$^{-3}$ for light silicate dust).
and $n$ is the refractive index, e.g. 1.5 for silicates.
The local ISM has $<A/L>=$1.8 mag/kpc, and we use 
standard values in \citet{Whittet1992book}:  
$\rho_{\rm dust} =2 \times 10^{-24}$ kg m$^{-3}$
and $\rho_{\rm gas}= 2 \times 10^{-22}$ kg m$^{-3}$, i.e. with 
a minimum gas-to-dust ratio of 100. 

In our case, the size of the cloud is $<$0.01pc (rather than typical ISM
path lengths of kpc), so the dust density we find is $\rho_{\rm dust}=2
\times 10^{-19} \mathrm{kg m^{-3}}$.  Thus $\rho_{\rm gas}$ is at least $2
\times 10^{-17} \mathrm{kg m^{-3}}$, and $M_{\rm cloud}=(4 \pi/3)
R_{\rm cloud}^3 \rho_{\rm gas}=0.002 M_{\odot}$ for $R_{\rm cloud}=0.01$pc. This
corresponds to a column density $N_H~10^{21}
\mathrm{cm^{-2}}$. Assuming the gas is cold, as required to prevent
sublimation of dust, this is sufficient to imprint a substantial,
narrow absorption feature onto the hydrogen emission lines, which is
not observed \cite[see e.g.][]{ReesNetzerFerland1989}.

Here we have assumed standard ISM dust to get an order of magnitude
estimate for the mass required to produce our total extinction;
however, as the \citet{Guo2016} Figure 6 extinction curve clearly
shows, we are not looking at standard ISM dust. Indeed, as those 
authors note, there are no identical extinction curves in the
literature, but the shape as a function of wavelength that suggests
extremely small scatterers. We note that although a very unusual cloud
might be able to explain our continuum extinction, given the observed
rest-frame $\approx$weeks-long timescales, the same scenario could not
explain the J2317+0005 event (despite the similar extinction curve).

\subsection{Scenario II: Accretion Disk Model}
Having discounted an obscuring event as the explanation for
J1100-0053, we turn to accretion disk models. We consider `cold'
accretion flows, described as optically thick, geometrically thin and
which drive relatively high mass accretion rates. They are `cold' in
the sense that the virial temperature of particles near the black hole
is low. Similarly, we characterize optically thin, geometrically thick
and low mass accretion rate flows as virially `hot' accretion
flows. \citet{YuanNarayan2014} present a detailed review on virially
hot accretion flows around black holes.

After giving our model set-up, we discuss whether J1100-0053 can be
described by a `hot' accretion flow, such as the advection-dominated
accretion flow. We then discuss our preferred `cold' accretion flow
model, but where the temperature of the accretion disk is perturbed by
propagating cooling and heating fronts in the inner parts ($\leq 1000
r_{g}$) of the accretion disk. Our disk remains virially cold
throughout this cycle.
 
We start with a multi-temperature blackbody (MTB) model, with a $L
\propto T^4$ dependence and a $T \propto r^{-3/4}$ relation. A thin
accretion disk has a negligible radial pressure gradient. Therefore,
at each radius $R$ the gas orbits at the Keplerian angular frequency,
$\Omega_{\rm K} = (GM/r^{3})^{1/2}$, where $M$ is the mass of the
central object, which possesses specific angular momentum $l=
\sqrt{GMr}$.

\begin{table*}
  \centering
  \begin{tabular}{l l l l l }
    \hline \hline 
    Model       &   Torque at ISCO  & AD suppression        &  Reddening law        &  Shown by \\
    \hline
    NZT            & \cmark        &  $\times$                        &   $\times$                            & Fig.~2 blue solid; Fig.~3 black solid \\
    ZT               &  $\times$    &  $\times$                        &   $\times$                            & Fig.~3 black dashed          \\
    ZT-abs              &  $\times$     &  $\times$                       &  Fit to \citet{Guo2016}   & Fig.~2 red long dashed   \\
    ZT-Spartfit  &  $\times$    &  90\%  inside 40$r_g$     &  $\times$                              & Fig.~2 red dotted              \\
    ZT-Snofit$^{*}$  &  $\times$     &  100\% inside 225$r_g$  &    $\times$            &  Fig.~3 black dotted         \\
    BB(7800K)$^{*}$  &   n/a             &   n/a                               &  n/a                         & Fig.~3 black dash-dot     \\
    \hline 
    \hline 
  \end{tabular}
  \caption{Salient details of the models presented in Figures~\ref{fig:J110057_spectra} and 
  \ref{fig:disk_suppression}. 
$^{*}$The ZT-Snofit and BB models are arbitrarily normalized to match the peak of the 2010 spectrum; the other models possess the same relative normalizations. 
We show the ZT-Snofit and BB models only to demonstrate the inability of grey-body absorption to fit the {\it shape} of the observed spectra even in the most extreme circumstances. We model the 
Guo et al. absorption as a piecewise 3-line fit to the extinction data in their Figure 6. 
The transmission fraction at that wavelength (in nm): 
$\lambda_{\rm break}=[714.0,435.0,370.0,285.0]$, ${\rm ext}_{\rm norm}=[1.0,0.87,0.70,0.25]$.
$\alpha$ is assumed to be 0.3 in all instances.
} 
 \label{tab:models}
\end{table*}

We present in Table~\ref{tab:models} the salient details of the models 
we now present and discuss below. 

\citet{Zimmerman2005} compare models with `zero' and `non-zero' torque
at the ISCO, and the impact on the temperature profile of the
corresponding accretion disk when the torque changes.  From
\citet{Zimmerman2005}, the zero torque (ZT) luminosity is given by
\begin{equation}
L_{\rm disk}   =  \frac{G M \dot{M}}  {2 r_{\rm in}}    = 73.9 \sigma\left ( \frac{T_{\rm max}}{f}  \right )^{4}  r^{2}_{\rm in} 
\end{equation}
and the standard, non-zero torque (NZT) luminosity is given by:
\begin{equation}
L_{\rm disk} = \frac{3 G M \dot{M}}  {2 r_{\rm in}}    = 12.6 \sigma\left ( \frac{T_{\rm max}}{f}  \right )^{4}  r^{2}_{\rm in} 
\end{equation} 
where $f$ is a spectral hardening factor, with a canonical blackbody
spectrum having $f=1$ and $\sigma$ is the Stefan-Boltzmann constant.
In zero-torque models, the temperature $T_{\rm ZT}$ goes to zero at
the inner edge of the disk (since the torque vanishes there) whereas
for a non-zero torque temperature profile, $T_{\rm NZT}$ reaches its
maximum value at the inner edge of the disk (where the torque is
maximal). Given a MTB model for disk emission, these differences at
small $r_{\rm g}$ translate to large differences in the SED.

Since we are concerned with inner parts of the disc, we should
consider relativistic effects. The ISCO probably lies at $r <
6GM/c^{2} = 3r_{\rm s}$ since quasar black holes generally have
prograde rotation \citep[e.g., ][]{Reynolds2014, Capellupo2016}. The
specific binding energy at the ISCO is therefore probably larger than
$1 - \sqrt{8/9} \approx 0.057$, so that even if there were no torque
at the ISCO, the ratio of the bolometric luminosity to the
monochromatic luminosity at any point in the rest-frame UV/optical
spectra could be larger than in a steadily accreting thin disc. We
note that since $l = \sqrt{GMr}$, we do not expect the ISCO
to change on human timescales.
 
As shown in Figure~\ref{fig:disk_suppression}, the SDSS spectrum from
2000 is well fit with a thin \citet{SS73} $\alpha$-disk and the NZT
condition.  However switching to just the zero-torque condition, while
surpressing the bluer disk emissivity, is not sufficient to explain
the 2010 spectrum. We note that we could have chosen a different
normalization in Figure 3, such that the ZT model would fit our 2000
spectrum; however, maintaining such a normalization would mean the
ZT-abs and ZT-Spartfit models would lie above our later spectra.

\subsubsection{Switching States to a RIAF/ADAFs:}
A possible explanation for the behaviour of J1100-0053 is that it
switches accretion modes, from a virially cold, high $\dot{M}$ flow to
a virially hot, low $\dot{M}$ flow. The latter could be either a
radiatively inefficient accretion flow \citep[RIAF; ][]{Narayan1998,
Quataert2001} or an advection-dominated accretion flow \citep[ADAF;
][and references therein]{YuanNarayan2014}.

There are examples of this type of behaviour in lower-luminosity
objects.  For example, \citet{Nemmen2006} successfully explain the SED
for the low-ionization nuclear emission-line region (LINER) of NGC
1097 with a model where the inner part of the flow is a virially hot
RIAF, and the outer part is a standard virially cold thin disk. The
broadband spectrum of NGC 1097 from \citet{Nemmen2006} initially
appears similar to the UV/optical 2010 spectrum of J1100-0053.  Figure
4 in \citet{Nemmen2006} shows the MTB-like model component from the
thin disk at $r>225r_{g}$ dramatically decreasing at $\sim$10$^{15}$Hz
($\sim$300nm). \citet{Nemmen2006} model the disk region interior to
this as a RIAF\footnote{A change to an ADAF is also possible in this
model.} at a power (in $\nu L_{\nu}$), an order of magnitude lower
than the MTB in the optical, but spanning from the X-ray to the
far-IR.

\begin{figure}
  \centering
  \includegraphics[width=8.7cm, trim=0.2cm 0.2cm 1.4cm 0.6cm, clip]
  {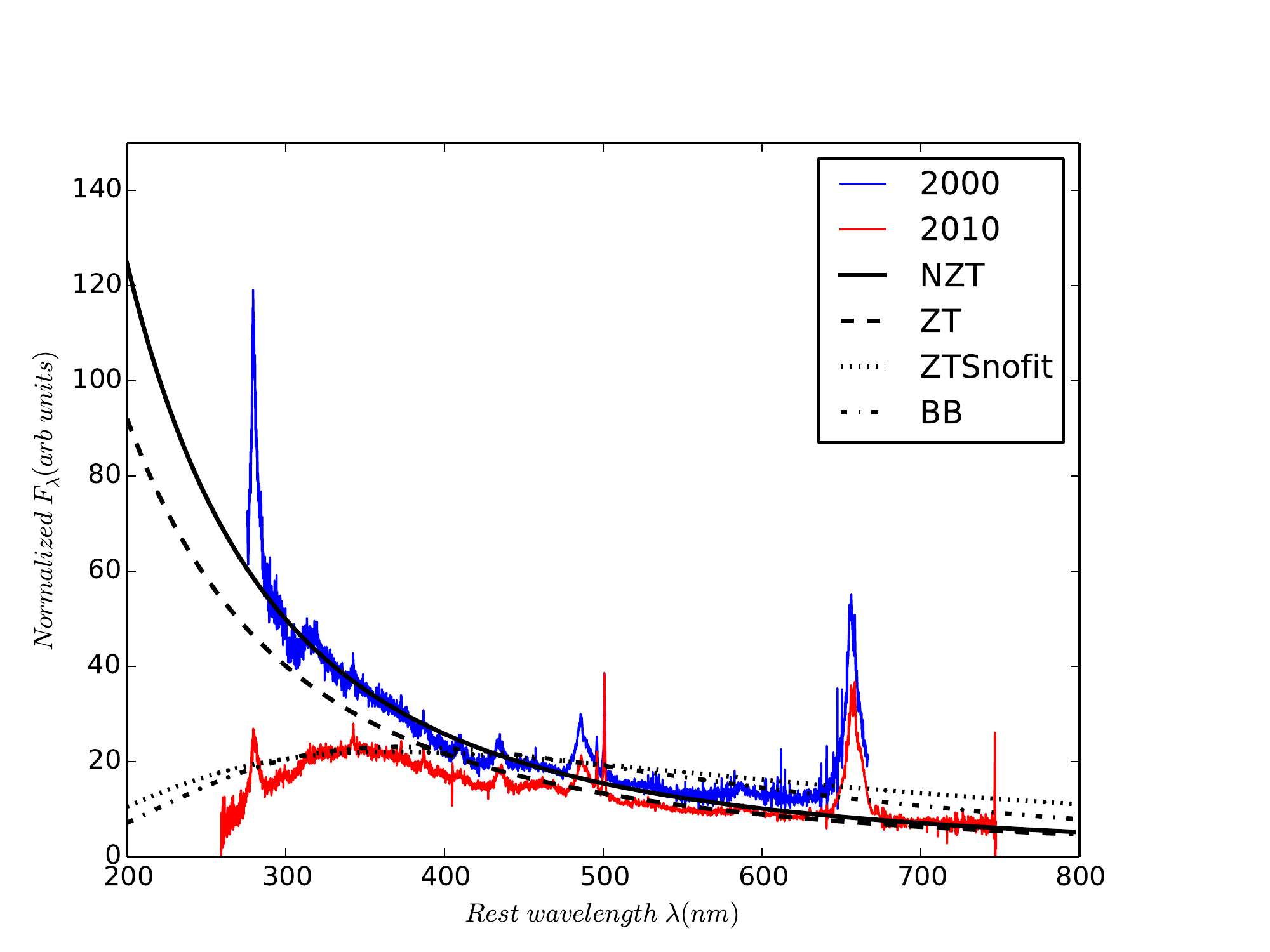}
  \vspace{-12pt}
  \caption[]{
    J1100-0053 data (blue line 2000 spectrum; red line 2010 spec-
    trum) with NZT model and 3 models providing poor fits. The solid black
    line shows non-zero torque at ISCO 
    \citep[following][]{Afshordi_Paczynski2003} and as in Figure 2, while the dashed black line shows a zero torque
    at the ISCO model. As noted in the text, a simple shift from a non-zero
    torque to a zero-torque solution is insufficient to reproduce the 2010
    spectrum. Indeed, no form of grey-body absorption can explain the 2010
    data, and we plot two example efforts -- the dotted line is the
    continuum from the same zero-torque model but with no emission from
    inside 225$r_{g}$ , normalized to the peak of the 2010 data. The dash-dot
    line is a single blackbody spectrum for $T=7800$K, again normalized to
    the 2010 peak. We note that though the temperature profile can be
    altered to produce a `turnover' in the neighborhood of 350nm, the
    shape of a blackbody is too broad. Worse, in this figure we have
    arbitrarily enhanced the bolometric luminosity expected from such a
    source, relative to the NZT and ZT models, as well as the models
    plotted in Figure 2 (which all possess the same normalization).
}
  \label{fig:disk_suppression}
\end{figure}
Can J1100-0053 switch states from a thin disk quasar to an ADAF at
small radii with the thin disk surviving at large radii?  Assuming the
transition happens due to a thermal instability in the inner disk on
the thermal timescale, and propagates outwards to radii
$\sim$225$r_{g}$ as in \citet{Nemmen2006}, we can parameterize the
front propagation time as
\begin{equation}
    t_{\rm front}  \sim  5 \; {\rm yr} \left(\frac{h/r}{0.1}\right)^{-1}
                                                           \left(\frac{\alpha}{0.3}\right)^{-1}  
                                                           \left(\frac{r}{225r_{g}}\right)^{3/2}  
                                                           \frac{r_{g}}{c}
\label{eqn:t_front}
\end{equation}
where we have had to assume a higher value of $\alpha \sim 0.3$
\citep{King2007} than typically assumed for thin ($h/r \ll 1$)
disks. This is plausible if there exists a very viscous disk and the
effect propagates outwards on a timescale of $\leq 5$ years from the
inner disk.

If the viscous disk switches to a RIAF at radii $<$225$r_{g}$, then the UV/optical emission should be suppressed by several orders of magnitude compared to a radiatively efficient thin disk \citep{Narayan1998, Abramowicz2002, Abramowicz2013}. However, if the thin disk emission is simply uniformly suppressed within $<225r_{g}$ by a large factor, we can not reproduce the shape of the J1100-0053 spectrum in 2010 (see Fig.~\ref{fig:disk_suppression}). Furthermore, in order to restore the thin disk in the 2016 observation, a thermal instability is required to occur at $\sim$225$r_{g}$ and a front to propagate inwards, collapsing the RIAF back to a thin disk.

Noting RIAFs/ADAFs exist at lower luminosity than for a classic thin
disk ($\epsilon \sim 0.005$ and $\epsilon \sim 0.1$, respectively, for
$L=\epsilon \dot{M} c^{2}$) it is unclear first what physical
processes would trigger the change of state to an ADAF and then cool
back down to a thin disk, and second, why such an instability would
occur at the thin disk/RIAF boundary in J1100-0053, whereas in NGC
1097 this interface appears to be stable. In any case, suppressing the
MTB temperature profile inside a radius of $225 r_{g}$ would lead to a
collapse in the total flux compared to unperturbed disk. These
scenarios are difficult to reconcile with our data.

\subsubsection{Changes at the ISCO and a cold absorbing phase}
An alternative explanation of our observations involves a triggering
event at the ISCO and an associated cold, absorbing or scattering
phase. Here we discuss the phenomenological requirements of this model
from the photometry in Fig.~\ref{fig:J110057_LC_CRTS}, the spectra in
Fig.~\ref{fig:J110057_spectra} and the line measurements from
Table~\ref{tab:Hbeta_details}. First we outline the model fits that are
required and then we attempt to construct a simple coherent
phenomenological picture.

The SDSS spectrum from 2000 can be well fit with a simple standard
thin accretion disk model with non-zero torque at the ISCO. The 2010
spectrum is relatively well fit with a thin disk model but now with
zero torque at the ISCO \emph{and} with a cold absorbing screen very
similar to that observed by \citet{Guo2016}. Both of these fits are
shown in Fig.~\ref{fig:disk_suppression}. From
Fig.~\ref{fig:J110057_LC_CRTS}, the PanSTARRS fluxes drop strongly in
2011, particularly in the green band, and remain low until about 2014
(MJD 56800), whereupon the DECaLS green and then red fluxes climb back
upwards. By 2017, the spectrum is well fit with a zero-torque disk
model with strong grey-body suppression of the flux to $10\%$ of an
unperturbed disk at radii interior to $40r_{g}$. This can be
interpreted as a modest change in the effective temperature in the
innermost disk at $\leq$40$r_{g}$. During this time the WISE W1 and W2
flux reaches a minimum around MJD 57000 (2015) and then climbs back
towards the values from 2010. The line measurements in
Table~\ref{tab:Hbeta_details} imply that the broad lines are stronger in
2010 than in 2017.

\begin{figure*}
  \includegraphics[width=15.4cm, trim=0.0cm 0.0cm 0.0cm 0.0cm, clip]
  {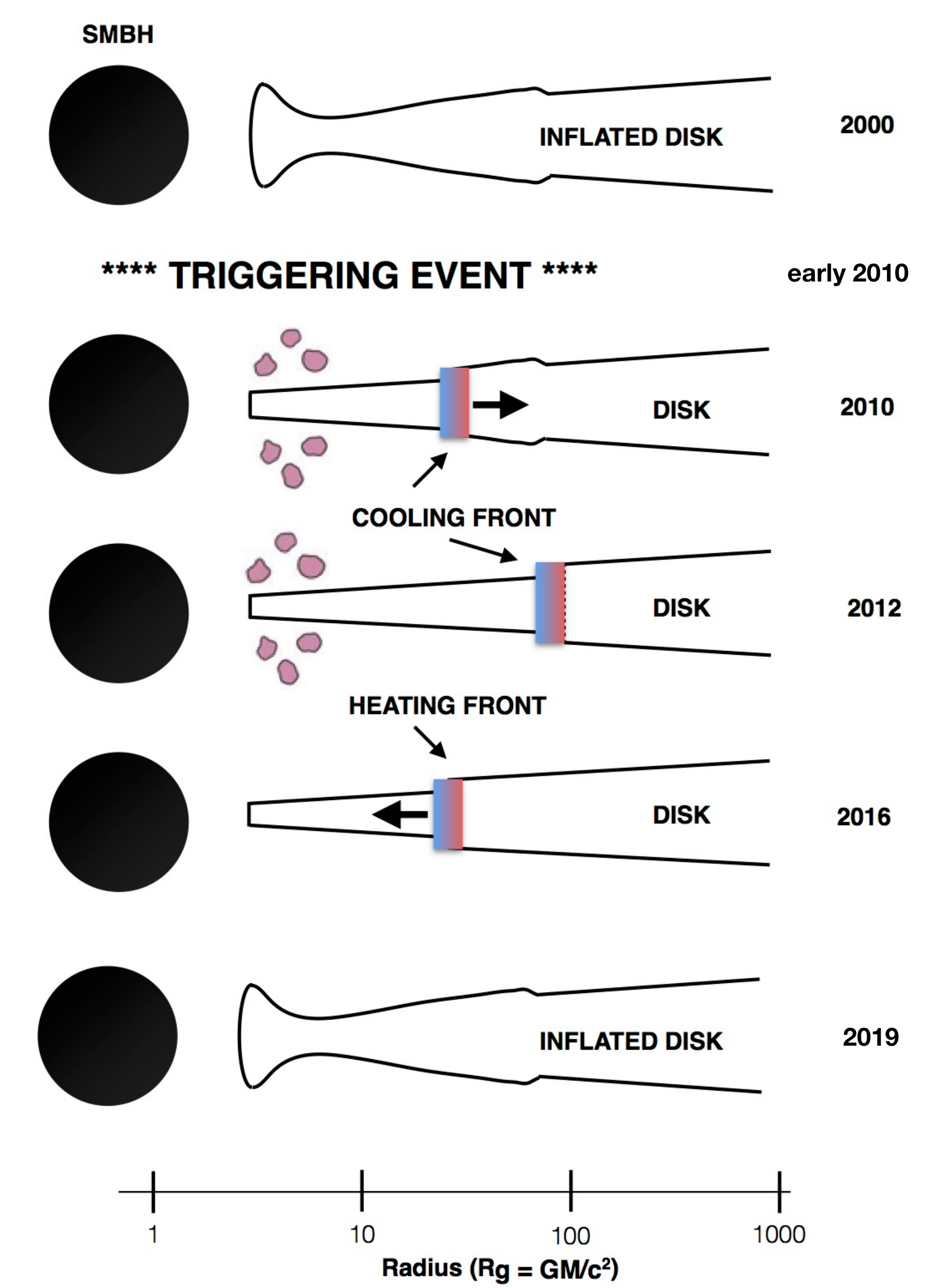}
  \centering
  \caption[]{
    Cartoon illustration of our model explaining the unusual spectral
evolution of J1100-0053. In 2000, corresponding to the SDSS spectral
epoch, the quasar has a standard inflated accretion disk, i.e., where
non-zero torque at the ISCO heats the inner radii of the accretion
disk, causing it to puff up \citep[e.g.,][]{Zimmerman2005}. Just
before the 2010 spectrum, a triggering event occurs that deflates the
inner disk, possibly due to a shift in the magnetic field
configuration leading to zero torque at the ISCO.  This event is
associated with some scattering/absorbing cold phase clouds, and
causes a cooling front to propagate outwards in the accretion disk,
traveling on the $t_{\rm front}$ time-scale \citep[see
also][]{Hameury2009}.  Circa 2014, a heating front travels radially
inwards, re-heating the inner accretion disk but on longer timescales,
due to the thinner disk. We predict that in the next year, the quasar
should roughly return to its initial state.}
  \label{fig:J110057_cartoon}
\end{figure*}
Putting all this together into a simple phenomenological model
suggests the following course of events, which we depict in cartoon
form in Fig.~\ref{fig:J110057_cartoon}.  Conditions change around 2010
at the ISCO and this change is associated with a cold absorbing or
scattering phase. Since the broad lines are relatively strong in 2010,
the change can not have occurred more than several months before the
2010 spectrum was taken. The triggering event may be a change in
$\dot{M}$ due to a stochastic variation in the mass supply or a local
change in $\alpha$. Since accretion disk luminosity is probably
powered by magnetized gas losing angular momentum, magnetized gas in
the plunging region might be expected to torque the ISCO gas
\citep[e.g., ][]{Gammie1999, Agol_Krolik2000}. If there is a change in
the magnetic field configuration around the SMBH, such that the torque
decreases to near zero at the ISCO, the temperature of the innermost disk will
drop dramatically \citep{Cao2003}.  We speculate that just such a
change occurred not long before the 2010 spectrum was taken.

Associated with a change at the ISCO, we require a cold, absorbing
and/or scattering phase. There are several possible sources of such a
cold phase. A scattering phase might arise from the collapse of a
corona. Or a cold phase could appear in the helical outflowing wind
which generates the innermost BLR. A dramatic change in disk surface
opacity might also be responsible, or some combination of all three of
these. So while in 2000 only one model component is needed to fit the
spectrum, namely a thin disk with non-zero torque at the ISCO, in 2010
two model components are required: a thin disk with zero-torque at the
ISCO \emph{and} a cold absorbing/scattering screen. The 2010 spectrum
can not be fit with any variety of greybody absorption, or
equivalently a simple MTB model with an alternate temperature
profile. It is, however, reasonably well fit using the wavelength
dependent absorption model of \citet{Guo2016}.

The photometric data in Fig.~\ref{fig:J110057_LC_CRTS} require a
further drop of around a magnitude in the PanSTARRS colours by
2011. We can achieve this by keeping the two-component model from
2010, but by adding a third component, a greybody suppression out to
$\sim$100$r_{g}$. We speculate that a cooling front propagates outwards
in the disk from the changes at the ISCO in 2010 so that regions of
the disk that contribute strongly to green and red emission are
suppressed by 2011. A cooling front propagates at speed $v_{\rm
front}=\alpha c_{\rm s}$ \citep{Hameury2009} so $\alpha \geq
0.1(c_{s}/10^{4}{\rm km s^{-1}})$ is required, but with $v_{\rm front}$
slowing the farther it travels since $c_{s}$ is expected to fall
rapidly with increasing radius \citep{Sirko_Goodman2003}.

By 2014, the PanSTARRS green and red fluxes begin recovery. This may
be associated with the inward propagation of a heating instability
\citep{Hameury2009}. For example, if the surface density, $\Sigma$,
reaches a critical value, a heating front can propagate back inwards,
analogous to the well-known accretion disk limit cycle mechanism in
models of dwarf novae outbursts \citep[e.g.,][]{Cannizzo1998}. The
returning heating front travels more slowly because the disk is colder
and thinner, and $t_{\rm front}$ is inversely proportional to
$h/r$. However, the heating from will re-inflate the disk as it
propagates inwards towards the SMBH. As a result, the recovery to 2010
fluxes in the PanSTARRS bands will take longer than the drop from
2010-2011.

While the cooling/heating has been going on in the accretion disk, the
initial dimming of the central ionizing flux will have an effect on
the distant (parsec-scale) torus. After a light travel time of $\sim$3
years the IR flux from the torus should drop \citep{Koshida2014,
Jun2015b} beginning in 2013, dropping to a minimum in 2014-15 (since
the disk is most suppressed in 2011 and the PanSTARRS band recover
slightly or remain flat in 2012-13). We speculate that the absorbing
screen has dissipated by this epoch in order to match the IR minimum
in 2015. This seems to be consistent with the behaviour of the source
in the WISE W1 and W2 bands. The cooling/heating front disk
propagation, the connection and origin of the BLR, and the effect on
the inner dust has also recently been explored in
\citet{Baskin_Laor2018}.

Assuming that nothing else drastic happens, our model predicts that
the heating front returns to the ISCO in late 2018 (around now). That
means the broad Balmer lines will reach a maximum a few months later
in early 2019, but the IR flux should not return to full flux until 2021. 

If the cold phase is associated with the condensation or disappearance
of part of the inner BLR, then the maximum EWs in 2019 may still be
less than the EWs measured in 2010. We also note that if J1100-0053 is
an ``anomalous H$\beta$ quasar'' as suggested by
\citet{Steinhardt_Silverman2013} and described above, then the
variable linewidths can be explained by the disk going into a low
state and the innermost, fastest, outflow disappears (due to,
e.g. reduced radiation pressure). However, we have not been able to
constrain the behaviour of H$\beta$ in J2317+0005 during the downturn
observed by \citep{Guo2016}. Therefore it is not clear whether there
is a general link between a broad H$\beta$ component and an apparent
state change in the CLQ disk. As such, we note this potentially
intriguing behaviour of the H$\beta$ emission line and look to
investigate further in the context of the ``changing-look'' phenomenon
in a future demographic study.

By 2017, the returning heating front has propagated back in to
$\sim$40$r_{g}$ and the cold phase has disappeared. So it seems the
cold phase is required between 2010-13, but then has fully disappeared
by 2017. This is an important constraint on the nature of the cold
phase which is an important but poorly understood component in our
phenomenological model. If, for example, the cold phase clumps on
size-scales of order $r_g$, with an overdensity of $10^{4}$ relative
to their hotter surroundings, and a relative velocity compared to the
hot phase on the order of the orbital velocity, then the clumps are
unstable to the Kelvin-Helmholtz instability
(eqn.~\ref{eqn:T_cloudcrushing}) on an approximate timescale of
\begin{equation}
t_{{\rm cc}} \approx 3{\rm mo} 
               \left( \frac{\rho_{\rm{cloud}}/\rho_{\rm{medium}}}{10^{4}}   \right)^{1/2}
               \left( \frac{r_{\rm cloud}}{r_{g}}\right) 
              \left( \frac{v_{\rm rel}}{10^{4}\rm{km s^{-1}}} \right)^{-1}.
\end{equation}
The important point here is that cold phase clouds are unstable to
collapse after a short time unless they are extremely over-dense with
very cold cores relative to the surrounding medium. The main coolants
at low temperatures are carbon and oxygen resonance lines and hydrogen
and helium from neutral phase material \citep[see e.g., Fig. 18 in
][]{Sutherland_Dopita1993}. The ionization energies for carbon and
oxygen are 11.26 and 13.61 eV, respectively, i.e., $\sim 100$nm, and
hence at wavelengths $<100$nm the disk opacity will increase
dramatically in edges. In the inner gas disk such edges must be
pressure, turbulence and Doppler broadened, but to depress the flux
out to $\sim$300nm would require associated velocities of
$\sim$0.9c. Thus, cooling absorption edges could explain the cold
absorbing phase only if substantial blanketing material is flung out
from near the event horizon, but this would not match the mild, longer
wavelength absorption required in \citet{Guo2016} or our
source. Rayleigh scattering might also help explain the steep cut-off
in the 2010 spectrum; we might expect an additional scattering layer
from slightly larger ions or neutral atoms in a cold or condensing
phase from the corona or the BLR. However, we can not take a strong
position on the identity of the absorber without more data, especially
time resolved spectra of an entire event such as this.

\section{Conclusions} 
By monitoring changing look quasars we introduce new tests of models
of accretion disk physics. We present the quasar J1100-0053 that was
catalogued in the Sloan Digital Sky Survey quasar survey, but
identified as an interesting due to its infrared photometric light
curve.

We have shown that a simple phenomenological model with a propagating
cooling front is capable of describing the gross spectral and temporal
variations in this CLQ. Our model makes a prediction for this source,
testable over the next few years and, if confirmed, implies that CLQs
as a class are driven by changes near the ISCO, close to the SMBH. The
discovery of J1100-0053 (and J2317+0005) are specific key examples of
time-domain astronomy and the resulting astrophysics to be
studied. However, even with the coverage from WISE, PanSTARRS, SDSS,
DECaLS and CRTS, we have a relatively sparse dataset which can not
tightly constrain our theoretical model.

The Zwicky Transient
Facility \citep[ZTF; ][]{Bellm2014} has very recently started and will
open a new data space with high cadence, multi-band photometric
monitoring. Along with ZTF in the very near future, the Large Synoptic
Survey Telescope \citep{Ivezic2008, LSST_ScienceBookV2} will allow
identification of the types of events such as J1100-0053 and
J2317+0005 \emph{while they are occurring}, allowing spectroscopic
monitoring. We will be able to see how long a UV collapse lasts and
closely follow its evolution.  Such data will stringently test models
of AGN disks at much higher fidelity than we are able to do with
current CLQ samples.

\smallskip
\smallskip
\noindent
{\bf Author Contributions.}   
N.P.R. led the project, developed and wrote the initial drafts of the manuscript and co-ordinated the team.
K.E.S.F. and B.K. orginated and developed the theoretical interpretation presented here. 
M.G. and D.S. were heavily responsible for the initial discussions and observations that were the genesis of
this project. A.M.M. produced the initial infrared variable quasar catalogs.  
D.S., M.G. and A.J.D. were part of the Palomar observing team.
N.P.R., A.M.M. and A.D. are part of the DECaLS Legacy Survey. 
R.A., A.D. and H.D.J. contributed to the manuscript.

\smallskip
\smallskip
\noindent
{\bf Availability of Data and computer analysis codes}. 
All materials, data, code and analysis algorithms are fully 
available at: 
\href{https://github.com/d80b2t/WISE\_LCs}{\tt https://github.com/d80b2t/WISE\_LCs}

\section*{Acknowledgements}
NPR acknowledges support from the STFC and the Ernest Rutherford
Fellowship scheme.  KESF \& BM are supported by NSF PAARE
AST-1153335. KESF \& BM thank CalTech/JPL for support during
sabbatical.  MF acknowledges support from NSF grants AST-1518308,
AST-1749235, AST-1413600 and NASA grant 16-ADAP16-0232.  RJA was
supported by FONDECYT grant number 1151408. AMM acknowledges support
from NASA ADAP grant NNH17AE75I. HDJ was supported by Basic Science
Research Program through the National Research Foundation of Korea
(NRF) funded by the Ministry of Education (NRF-2017R1A6A3A04005158).

We thank David J. Schlegel for quality checks on the BOSS data, and
Chris Done for invigorating discussions at the concept and conclusion
of this work. We also thank Giorgio Calderone for discussions on 
QSFit.  

This publication makes use of data products from the Wide-field
Infrared Survey Explorer, which is a joint project of the University
of California, Los Angeles, and the Jet Propulsion
Laboratory/California Institute of Technology, and NEOWISE, which is a
project of the Jet Propulsion Laboratory/California Institute of
Technology. WISE and NEOWISE are funded by the National Aeronautics
and Space Administration.

This research has made use of the NASA/IPAC Extragalactic Database
(NED) which is operated by the Jet Propulsion Laboratory, California
Institute of Technology, under contract with the National Aeronautics
and Space Administration.

This research has made use of data obtained from the SuperCOSMOS
Science Archive, prepared and hosted by the Wide Field Astronomy Unit,
Institute for Astronomy, University of Edinburgh, which is funded by
the UK Science and Technology Facilities Council.

The GALEX GR6/7 Data Release hosted at
\href{http://galex.stsci.edu/GR6/}{http://galex.stsci.edu/GR6/} was
used. These data were obtained from the Mikulski Archive for Space
Telescopes (MAST). STScI is operated by the Association of
Universities for Research in Astronomy, Inc., under NASA contract
NAS5-26555. Support for MAST for non-HST data is provided by the NASA
Office of Space Science via grant NNX09AF08G and by other grants and
contracts.

Funding for SDSS-III has been provided by the Alfred P. Sloan
Foundation, the Participating Institutions, the National Science
Foundation, and the U.S. Department of Energy Office of Science. The
SDSS-III web site is
\href{http://www.sdss3.org/}{http://www.sdss3.org/}.
SDSS-III is managed by the Astrophysical Research Consortium for the
Participating Institutions of the SDSS-III Collaboration including the
University of Arizona, the Brazilian Participation Group, Brookhaven
National Laboratory, Carnegie Mellon University, University of
Florida, the French Participation Group, the German Participation
Group, Harvard University, the Instituto de Astrofisica de Canarias,
the Michigan State/Notre Dame/JINA Participation Group, Johns Hopkins
University, Lawrence Berkeley National Laboratory, Max Planck
Institute for Astrophysics, Max Planck Institute for Extraterrestrial
Physics, New Mexico State University, New York University, Ohio State
University, Pennsylvania State University, University of Portsmouth,
Princeton University, the Spanish Participation Group, University of
Tokyo, University of Utah, Vanderbilt University, University of
Virginia, University of Washington, and Yale University.

\bibliographystyle{mnras}
\bibliography{/cos_pc19a_npr/LaTeX/tester_mnras}

\bsp	
\label{lastpage}
\end{document}